\documentclass[prc,twocolumn]{revtex4-1}
\usepackage{amsmath,graphicx}
\begin{document}
\title{Collective flow and long-range correlations in relativistic heavy ion collisions}
\author{Matthew Luzum}
\affiliation{
CEA, Institut de physique th\'eorique de Saclay (IPhT), F-91191
Gif-sur-Yvette, France}
\begin{abstract}
Making use of recently released data on dihadron correlations by the STAR collaboration, I analyze the long-range (``ridge-like'') part of these data and show that the dependence on both transverse momentum as well as orientation with respect to the event plane are consistent with correlations expected from only collective flow.  In combination with previously analyzed centrality-dependent data, they provide strong evidence that only collective flow effects are present at large relative pseudorapidy.
 In contrast,  by analyzing a ``background subtracted'' signal, the authors presenting the new data concluded that the ridge-like part of the measured correlation could not in fact be entirely generated from collective flow of the medium.   I explain the discrepancy and illustrate some pitfalls of using the ZYAM prescription to remove flow background.  
\end{abstract}
\maketitle
\section{Introduction}
Two-particle correlations in relativistic heavy ion collisions --- the probability of seeing a pair of particles with relative azimuth $\Delta\phi \equiv (\phi_1 - \phi_2)$ and relative pseudorapidity $\Delta\eta \equiv (\eta_1 - \eta_2$) in a given collision event --- display unique features not seen in other types of collisions such as p-p or d-Au.  In particular are significant long-range correlations extending to large $\Delta\eta$, which often show interesting ridge and shoulder structures after a model-dependent subtraction of elliptic flow \cite{:2008cqb, :2009qa, Alver:2009id}.

Recently it was proposed that this long-range part of the correlation could be entirely explained by collective flow effects \cite{Alver:2010gr, Sorensen:2010zq}.  Central to this idea was the fact that, due to event-by-event fluctuations, there should exist not only elliptic flow, but also ``triangular flow'', which should add a non-negligible contribution to these data.  Although triangular flow ($v_3$) has not yet been directly measured at RHIC, there were hints from a transport model, and it was later shown \cite{Alver:2010dn} from viscous hydrodynamic calculations, that the centrality dependence as well as the size 
(depending on viscosity) 
of the third Fourier component of the correlation $V_{3\Delta} \equiv \langle \cos(3\Delta\phi) \rangle$ 
does indeed quantitatively match that expected as arising simply from triangular flow.  Thus, the measured dihadron correlation at  large $\Delta\eta$, consists almost entirely of the lowest few Fourier components, each of which can be quantitatively understood as coming from collective flow (plus global momentum conservation) --- at least for the data analyzed, which had transverse momentum triggers of $p_t = 2.5$ GeV and lower.

Even more recently, the STAR collaboration has released data 
from 200 A*GeV Au-Au collisions
at RHIC, where a  trigger particle (with a $p_t$ of 3--4 GeV or 4--6 GeV) was restricted to be at a fixed angle with respect to the measured event plane, and its correlation with associated hadrons in various $p_t$ bins was measured \cite{Agakishiev:2010ur}.  The data were then separated into ``ridge-like'' and ``jet-like'' components, where the ridge-like correlation was defined by a projection on $|\Delta\eta|>0.7$, while the jet-like correlation was defined by taking the total short-range correlation at $|\Delta\eta|<0.7$ and subtracting the ridge-like correlation.

These new data provide an opportunity to test the idea that the long-range ridge-like correlation may be generated exclusively by collective flow effects.

In Ref.~\cite{Agakishiev:2010ur}, the ridge-like correlation was analyzed using the fairly common Zero-Yield-At-Minimum (ZYAM) prescription \cite{Adler:2005ee} to ``subtract'' the elliptic flow signal which dominates the unsubtracted data, in order to study non-flow behavior of the system.  
These flow-subtracted data show a dependence on the angle between the trigger particle and the event plane, and this was taken as evidence that collective flow alone could not explain this signal since triangular flow should be uncorrelated with the event plane.

In this article I show that this conclusion is not correct.  As was seen in previously analyzed data, the long-range ``ridge-like'' correlation is entirely consistent with what is expected purely from collective flow, and rather than contradicting this idea, the new data actually provide strong evidence.  I then illustrate how the misleading background-subtracted signal seen by STAR can be generated by the use of the ZYAM prescription in a situation where the assumptions of such a subtraction procedure do not hold.

%
%
%
%
\begin{figure}
\includegraphics[width=\linewidth]{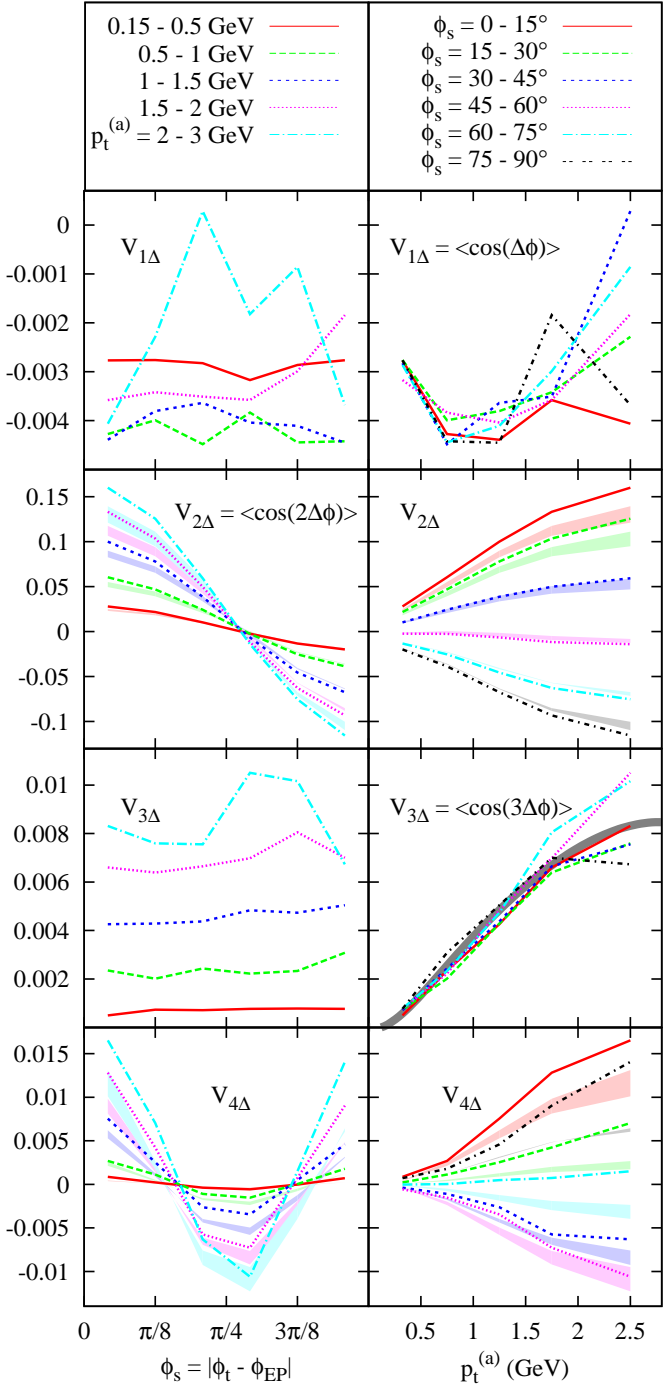}
\caption{(Color online) First 4 harmonics $V_{n\Delta} \equiv \langle \cos (n\Delta\phi) \rangle$ of dihadron correlations for particle pairs with relative angle $\Delta\phi = \phi_t - \phi_a$, and relative pseudorapidity $|\Delta\eta|>0.7$ \cite{Agakishiev:2010ur}.  On the left $V_{n\Delta}$ is plotted as a function of angle $\phi_s$ between a 3--4 GeV trigger particle and the event plane, and on the right as a function of transverse momentum of the associated particle, $p_t^{(a)}$.  The even harmonics also include bands for the estimated flow background
\cite{Agakishiev:2010ur}. Points are placed at the midpoint of each bin.  The grey curve of $V_{3\Delta}$ versus $p_t$ shows the $p_t$ dependence of triangular flow from viscous hydrodynamics (see text for details).  Data for 4--6 GeV trigger particles show the same trends, but with larger statistical fluctuations. }
\label{fig:vn}
\end{figure}
\section{Fourier decomposition of long-range correlations}
\label{flow}
It is useful to do a Fourier decomposition of unsubtracted dihadron correlation data at large relative pseudorapidity.  These data contain only long-range correlations, which stem from early times in the collision evolution \cite{Dumitru:2008wn}, and it has been argued \cite{Alver:2010gr} that they may contain significant contributions only from the collective behavior of the system (along with a trivial correlation from global momentum conservation).  The recent data from STAR 
present an opportunity to further test this idea by adding information about orientation with respect to the event plane as well as $p_t$ dependence, which complements the previous analyses \cite{Alver:2010gr, Alver:2010dn} of centrality dependent data from PHOBOS \cite{:2008gk, Alver:2009id} and STAR \cite{Abelev:2008un}.  

A Fourier analysis was done of the ridge-like component of the recent data (i.e., correlations for $|\Delta\eta|>0.7$) \cite{Agakishiev:2010ur}.  At this 20--60\% centrality range, the correlation is dominated by the second Fourier component $V_{2\Delta} \equiv \langle \cos(2\Delta\phi) \rangle$ (unless the angle of the trigger particle with respect to the event plane is close to $\pi/4$).  
This is followed by the remainder of the first four harmonics ($V_{4\Delta}$, $V_{3\Delta}$, and $V_{1\Delta}$) while all higher moments are significantly smaller --- the data in every bin can be very well approximated by the first four Fourier coefficients,
\begin{align}
\label{dndphi}
\frac {dN} {d\Delta\phi} \simeq \frac N {2\pi} [ & 1 + 2  V_{1\Delta} \cos (\Delta\phi) + 2  V_{2\Delta} \cos (2\Delta\phi)  \nonumber \\
& + 2  V_{3\Delta} \cos (3\Delta\phi) + 2  V_{4\Delta} \cos (4\Delta\phi) ] .
\end{align}

In general, collective flow introduces no direct correlation between pairs of particles.  In the absence of non-flow correlations, the two-particle correlation is determined solely by the one-particle distribution, and so in a given collision event the flow contribution to each harmonic factorizes:
\begin{align}
V_{n\Delta} \equiv& \langle \cos(n\Delta\phi) \rangle = {\rm Re} \langle e^{in(\phi_a - \phi_t)} \rangle \nonumber\\
 =& {\rm Re} \left(\langle e^{in(\phi_a - \psi_n)} \rangle \langle e^{-in(\phi_t-\psi_n)} \rangle \right) \nonumber\\
 =& \langle \cos(n\phi_a - n\psi_n) \rangle \langle \cos(n\phi_t - n\psi_n) \rangle 
\end{align}
Here $\phi_a$ and $\phi_t$ are the angles of the associated and trigger particle, respectively.  $\psi_n$ is an angle that can be determined separately for each $n$ in a given event in the same way as the event plane angle for elliptic flow, $\psi_2 = \psi_{EP}$.  Due to event-by-event fluctuations, these angles are not in general equal to each other or to the reaction plane that is defined by the beam and impact parameter.   Note also that when the correlation is averaged over many events, the average of the product is sensitive to the magnitude of these fluctuations.

Specifically how these two factors depend on independently measured quantities (e.g., $v_2$, etc. depends on how the pairs are selected and averaged.  This is discussed for each individual Fourier harmonic in the following sections:  
\subsection{The even harmonics: elliptic and quadrangular flow}
The dominant harmonic, $V_{2\Delta}$, is understood as coming largely from elliptic flow.  The flow contribution is \cite{Adcox:2002ms}:
\begin{align}
\label{v2}
V_{2\Delta} \equiv& \langle \cos(2\Delta\phi) \rangle = v_2^{(a)} v_2^{(t, R)}.
 %
\end{align}
%
Here $v_2^{(a)} = \langle \cos(2\phi_a-2\psi_{EP})\rangle$ is the elliptic flow of particles in the same bin as the associated particle, while $v_2^{(t,R)} \sim \cos(2\phi_s)$ since the trigger particle is fixed to be at a particular angle $\phi_s = |\phi_t - \psi_{EP}|$ with respect to the event plane, but there are corrections due to the finite angular bin width and event plane resolution \cite{Bielcikova:2003ku}.  These corrections depend on $v_2$ and $v_4$ of particles in the same $p_t$ bin as the trigger particle (see Eq.~(4) of Ref.~\cite{Agakishiev:2010ur} for the precise functional form of $v_2^{(t,R)}$ used by STAR).   

Figure~\ref{fig:vn} (row 2) shows the extracted $V_{2\Delta}$ along with the values of $v_2^{(a)} v_2^{(t, R)}$ reported by STAR (with bands representing their estimated systematic uncertainty).
By comparing the extracted values and estimated background, one can see that the data show the behavior expected from collective flow both as a function of $p_t$ and trigger angle.  The only difference is that the magnitude of the estimated flow contribution is consistently smaller than the extracted Fourier component --- the implied non-flow signal has roughly the same $p_t$ dependence as flow, and contributes a correlation that varies monotonically as the orientation goes from in-plane to out-of-plane; i.e., it has the same behavior as elliptic flow.

So either the background estimation is correct and the non-flow contribution to $\langle \cos2(\Delta\phi) \rangle$ has the same properties as flow, or the flow background is underestimated.  In fact, the latter is quite likely when the method of obtaining the estimation is analyzed.  

\label{v2meas}
The lower bound of the $v_2$ systematic uncertainty band for both the associated and trigger particle are given by measurements of the four particle cumulant elliptic flow $v_2\{4\}$.  This measurement significantly reduces the effects of non-flow \cite{Borghini:2001vi}, but also has a negative contribution from elliptic flow fluctuations \cite{Ollitrault:2009ie}.  Because of the imperfect event plane resolution, however, the correlation $V_{2\Delta}$ here has a positive contribution from elliptic flow fluctuations, 
and so the lower edge of the background uncertainty band is certainly smaller than the actual flow contribution.

The upper bound of the uncertainty band is obtained by using an event plane $v_2\{EP\}$ for the trigger particle and an away-side two-particle cumulant $v_2\{2,AS\}$ for the associated particle, though the dominant contribution is from the latter (recall that $v_2$ of the trigger particle only comes in as a small correction due to finite bin width and event plane resolution). 

The away-side two-particle cumulant is obtained in essentially the same way as one would obtain the second Fourier component of a two-particle correlation, except the correlation is only integrated over the away side ($|\Delta\phi|>\frac \pi 2$) instead of the entire azimuth.  This is presumably done to remove the non-flow contribution of the jet peak at small $\Delta\phi$ (and small $\Delta\eta$).  However, as a consequence, one is no longer actually picking out the second harmonic.  As mentioned in Ref.~\cite{Agakishiev:2010ur}, the negative dipole $V_{1\Delta}$ tends to increase $v_2\{2,AS\}$.  However, as can be seen by integrating Eq.~\ref{dndphi} over $|\Delta\phi|>\frac \pi 2$, a third harmonic $V_{3\Delta}$ will \textit{decrease} the value by an even larger amount (as long as it is at least as large as $\frac 5 9 V_{1\Delta}$, which is expected and also coincides with observation).

It is quite likely, then, that even the upper uncertainty bound of the estimated background is smaller than the actual elliptic flow contribution.

According to Ref~\cite{Agakishiev:2010ur}, $V_{4\Delta}$ has a similar contribution from collective flow:
\begin{equation}
\label{v4}
V_{4\Delta} \equiv \langle \cos(4\Delta\phi) \rangle \simeq v_4^{(a)} v_4^{(t, R)},
\end{equation}
where $v_4^{(a)} = \langle \cos(4\phi_a-4\psi_{EP})\rangle$ is the associated quadrangular flow measured with respect to the event plane and $v_4^{(t,R)}$ behaves roughly as $\cos(4\phi_s)$ (see Eq.~(5) of Ref.~\cite{Agakishiev:2010ur}).  Actually, this assumes that in each event the 4th harmonic of the single particle distribution is centered about the same angle as the 2nd harmonic ($\psi_4 = \psi_{EP}$).  In reality, the particle distributions define an angle $\psi_4$ that, like the event plane, fluctuates around the reaction plane event-by-event.  The actual contribution can be written as Eq.~\eqref{v4} plus a somewhat complicated additive correction that depends on event-by-event fluctuations.  So far, $v_4$ has only been measured with respect to the event plane, and not with respect to its own angle $\psi_4$, and a more complete analysis is left to future work, but here it is important to note only that this correction is expected to be relatively small, but always positive.

Figure~\ref{fig:vn} (row 4) shows $V_{4\Delta}$ as well as the expected background $v_4^{(a)} v_4^{(t, R)}$ as used in Ref.~\cite{Agakishiev:2010ur}.  One can see by comparing the extracted curves to the bands representing the expected background that it behaves as expected as a function of $p_t$ and angle, 
and actually shows evidence of $\psi_4$ fluctuations, since the extracted curves are always slightly above the estimated background.   Any non-flow contribution would need to have the same angular and $p_t$ dependence as flow to see these observed trends, or it would need to be very small and mimic the effects of $\psi_4$ fluctuations. Note that this means any non-flow signal must have an angular dependence for the 4th Fourier harmonic that is quite different from that of its 2nd harmonic.

Thus, the observed even harmonics are consistent with those expected from flow alone, according to every possible test criterion.
\subsection{The odd harmonics:  triangular flow, momentum conservation, and directed flow}
The third component, $V_{3\Delta}$, will have a contribution from collective flow of the form \cite{Alver:2010gr, Alver:2010dn}
\begin{equation}
\label{v3}
V_{3\Delta} \equiv \langle \cos(3\Delta\phi) \rangle \simeq v_3^{(a)} v_3^{(t)},
\end{equation}
where $v_3^{(a)} = \langle \cos(3\phi_a-3\psi_3)\rangle$ is the triangular flow of the associated particle, and $v_3^{(t)}$ is similarly
 the triangular flow for the trigger particle.  These are flow parameters that could be measured with respect to an event-by-event plane of triangularity $\psi_3$ that is expected to be largely uncorrelated with the event plane~\cite{Alver:2010gr, Staig:2010pn, Nagle:2010zk}, 
 and so this contribution should not depend on the event plane angle.  

Indeed, the data show no systematic dependence on event plane orientation (see Fig.~\ref{fig:vn}, row 3).  In addition, although no direct measurement of $v_3$ has yet been made, both the size and $p_t$ dependence are consistent with hydrodynamic predictions 
(the solid grey curve in Fig.~\ref{fig:vn} shows the $p_t$-differential $v_3$ for a viscous hydrodynamic calculation of a 30--35\% central collision with $\eta/s = \frac 1 {4\pi}$ and Glauber initial conditions from Ref.~\cite{Alver:2010dn}, multiplied by 0.09, representing the estimated $v_3^{(t)}$ of the trigger particle at 3--4 GeV).  Note that the centrality dependence of this correlation has also been shown to follow that calculated in hydrodynamic models \cite{Alver:2010dn}.  

Thus, if there is a non-flow signal, it must have a $\langle \cos3(\Delta\phi) \rangle$ component that is independent of angle, and has a $p_t$ dependence that is not too different from triangular flow.

The first harmonic, $V_{1\Delta}$, has a contribution from global momentum conservation shared between a finite multiplicity \cite{Borghini:2000cm}.  Although this is not, strictly speaking, a flow correlation, it will be present in any finite-multiplicity system, and so is not an interesting non-flow signal.  It does not have any dependence on the angle with respect to the event plane but should be negative with a linear dependence on $p_t$ of the associated particle \cite{Borghini:2000cm}. 

In addition, there is a possible contribution from collective flow.
The directed flow that has been measured at RHIC with respect to the impact parameter is very small everywhere within the acceptance of the STAR TPC \cite{Abelev:2008jga}. 
  Thus, one might expect no contribution from collective flow.  However, there can also be a $v_1$ that is generated event-by-event due to fluctuations in the initial geometry \cite{Teaney:2010vd}.   This contribution is expected to be approximately independent of rapidity, and therefore cancels in the existing directed flow measurements, which are odd in rapidity by construction \cite{Ollitrault:1997vz, Danielewicz:1985hn}.  However, it does not cancel in this dihadron correlation, and the effect is given by \cite{Teaney:2010vd, Wang:1991qh}
\begin{equation}
\label{v1}
V_{1\Delta} = \langle \cos(\Delta\phi) \rangle \sim v_1^{(a)} v_1^{(t)} ,
\end{equation}
where $v_1 = \langle \cos(\phi-\psi_1)\rangle$ is the directed flow that can be defined with respect to an event-by-event angle $\psi_1$. 
Generically it should be negative at low $p_t$ and positive at high $p_t$ \cite{Teaney:2010vd}, with a zero crossing at $\sim$800 MeV such that the net momentum in the direction of $\psi_1$ is approximately zero.  As mentioned, it is not directly correlated with the event plane, so, like the correlation from momentum conservation, it will also have no angular dependence.

%
%

As expected, the dipole harmonic indeed has no systematic dependence on trigger particle angle (see  Fig.~\ref{fig:vn}, row 1).  It turns out that the $p_t$ dependence is consistent with a contribution from both momentum conservation and flow, of roughly the expected size \cite{Teaney:2010vd}, although a more quantitative analysis of this $v_1$ and a proposal for how to measure it directly will be presented separately \cite{Luzum:2010fb}.  

 The most important point here is the lack of angular dependence, which is consistent with both effects, whatever their size or dependence on $p_t$.

It should also be noted that there is no significant contribution expected from collective flow for the 5th harmonic and higher, since the anisotropy coefficients $v_5$ and above were found to have a very small hydrodynamic response and very large viscous suppression \cite{Alver:2010dn}.  The absence of higher Fourier harmonics provide another piece of evidence in favor of the absence of non-flow signals.

All odd harmonics, then, have no dependence on the angle with the event plane.  This fact as well as the dependence on $p_t$ is entirely consistent with the properties expected from only flow and momentum conservation.  Any non-flow contribution would have to have these same properties to be consistent with data.
\subsection{Dependence on trigger $p_t$}
The plots in Fig.~\ref{fig:vn} were all made from the set of data where all trigger particles had transverse momentum in the range 3--4 GeV, since the statistical uncertainty is smaller than the other set of data with trigger $p_t$ of 4--6 GeV, and one can more accurately assess trends in the data.  Comparing the two data sets, however, provides another way to test whether there is a contribution other than collective flow.  The flow contribution of both sets of data should have the same shape as a function of $p_t$, with only a difference in normalization from $v_n^{(t)}$ of the trigger particle.   If there is a non-flow signal that becomes more prominent above 4 GeV, it should be apparent.

Remarkably, the data match expectations from flow.  Not only do the two sets share the same features as a function of angle, but they also have the same $p_t$ dependence.  Further, the difference in absolute normalization of the harmonics can be used to infer the large $p_t$ dependence of flow coefficients that have not yet been measured --- 
 comparing normalizations implies that, like $v_2$, $v_3$ also appears to turn over such that the value for particles with $p_t $= 4--6 GeV is slightly ($\sim 10\%$) lower than at 3--4 GeV.  This is both an interesting and reasonable result.
\subsection{Summary of Fourier analysis}
In summary, to be consistent with data, a non-flow signal in the analyzed ridge-like dihadron correlation would need to have odd Fourier components with no dependence on the angle $\phi_s$ with respect to the event plane, a second harmonic with a monotonically decreasing angular dependence, 
and a fourth harmonic that decreases and then increases with angle.  It would also need to have a dependence on $p_t$ of both the trigger and associated particle, in each harmonic, that is the same as that expected for flow.  In other words, it would have to have all the properties of flow.

A more sensible explanation is that the long-range correlation data in fact contain measurable contributions only from collective flow.
\section{ZYAM background subtraction and spurious signals}
The authors of Ref.~\cite{Agakishiev:2010ur} came to a different conclusion concerning the ridge-like correlation.  It is useful to understand how they came to this conclusion, and show how the ZYAM subtraction procedure can give misleading results.

As the name implies, the Zero-Yield-At-Minimum prescription makes specific assumptions about dihadron correlations, and in a case where such assumptions are erroneous, the subtracted signal provides questionable information at best.

Specifically, one must first assume precise knowledge of the flow background --- often this is assumed to be just a contribution from measured elliptic flow, but sometimes this is improved by including quadrangular flow, as in Ref.~\cite{Agakishiev:2010ur}.  Then, in the presence of non-flow, the overall measured correlation is of the form \cite{Adler:2005ee}
\begin{align}
\label{ZYAM}
\frac {dN} {d\Delta\phi} =& B \left[ {\rm F(\Delta\phi)}\right] + {\rm NF}(\Delta\phi) \nonumber \\
 =& B \left[ 1 + 2 v_2^{(a)} v_2^{(t,R)} \cos(2\Delta\phi) + 2 v_4^{(a)} v_4^{(t,R)} \cos(4\Delta\phi)\right] \nonumber\\
 &\ \ \ \   + {\rm NF(\Delta\phi)}
\end{align}

Here, F is the assumed flow background and NF is the remaining non-flow signal.  The flow background is typically assumed to be as in Eq.~\eqref{v2}, and sometimes including Eq.~\eqref{v4}.  In principle, the input flow coefficients (e.g., $v_n$), should come from an independent measurement that does not contain a contribution from non-flow correlations, but still has the same dependence on flow fluctuations.  Such a measurement does not exist, and it should be noted that it can be a very non-trivial exercise to determine the correct input values precisely enough to reliably extract a small non-flow correlation (see, e.g., the discussion of the $v_2$ estimation in Sec.~\ref{v2meas}).  Also note the complete lack of data for odd flow coefficients to date, which precludes their use in such subtraction schemes.

Since the non-flow contribution is unknown, the overall normalization, $B$, is now a free parameter.  To extract the desired non-flow signal, one fixes $B$ by assuming the non-flow contribution NF is always positive except at one or more minima, where it is zero (i.e., zero yield at minimum):

\begin{align}
{\rm NF}(\Delta\phi) = \frac {dN} {d\Delta\phi}  - B [1&  + 2 v_2^{(a)} v_2^{(t,R)} \cos(2\Delta\phi)  \nonumber \\
&+ 2 v_4^{(a)} v_4^{(t,R)} \cos(4\Delta\phi)] ,
\end{align}
\begin{align}
{\rm NF}(\Delta\phi_{\min}) = {\rm NF'}(\Delta\phi_{\min})  = 0 , \\
B = \left(\frac {dN} {d\Delta\phi}\right)_{\Delta\phi_{\min}} \frac 1 {{\rm F}(\Delta\phi_{\min})}.
\end{align}

Since dihadron correlations in, e.g.,  proton-proton and deuteron-gold collisions do typically have close to zero yield at their minimum, it may seem reasonable to suspect that the non-flow contribution to heavy ion collisions may also have this form.  However, there is no particular reason that this should be true.  

More importantly,  even if it is true, the ZYAM prescription can only extract this signal if the background is known to very good accuracy.  Due to event-by-event fluctuations, however, this is fraught with difficulty.  As mentioned, it is not possible to independently measure the input flow coefficients $v_n$ in such a way that removes non-flow correlations but still has the same contribution from flow fluctuations.  Thus, there will always be model dependence and uncertainty in these parameters that, small as they may be, compete with the level of precision needed to extract the desired signal of interest.  If any of these coefficients are not correctly estimated with enough precision (and certainly if the odd harmonics are not considered) the ZYAM procedure will not actually ``subtract'' even the assumed background, and the result will have significant contribution from flow.  This will be true even if the zero-yield-at-minimum assumption itself is valid.

\begin{figure}
\includegraphics[width=\linewidth]{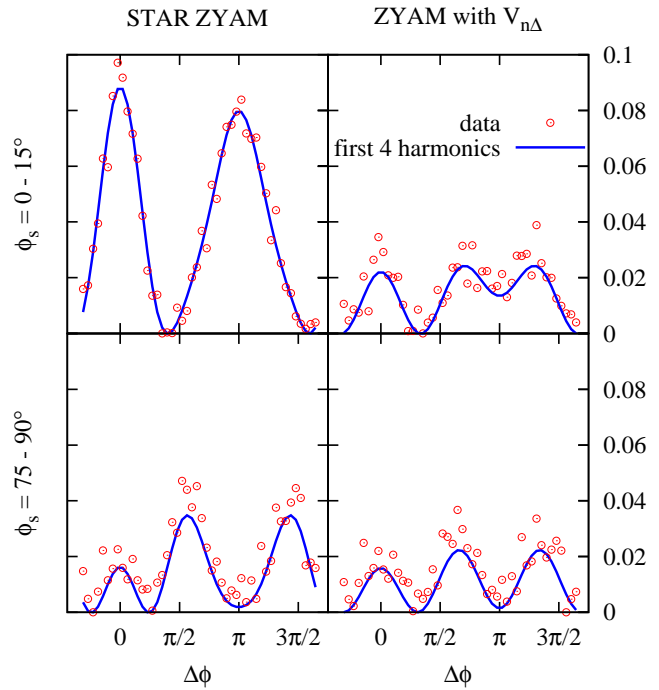}
\caption{(Color online) 
Background subtracted correlation $\frac 1 {N_{trig}} \frac {dN} {d\Delta\phi} \equiv {\rm NF}(\Delta\phi)$ for trigger $p_t  = $ 3--4 GeV and associated $p_t = $ 1.5--2 GeV.  Circles are measured data, while solid lines represent a sum of the first four harmonics extracted from the unsubtracted data.   
The left column represents ZYAM subtraction with the estimated background from Ref.~\cite{Agakishiev:2010ur}, while the right column uses the extracted harmonics 
$v_2^{(a)} v_2^{(t,R)}\equiv V_{2\Delta}$, $v_4^{(a)} v_4^{(t,R)}\equiv V_{4\Delta}$.
 The upper row is for an in-plane trigger particle, while the lower row is out-of-plane.}
\label{fig:ZYAM}
\end{figure}

This is illustrated in Fig.~\ref{fig:ZYAM}.  The circles in the left column represent ZYAM-subtracted data as in Ref.~\cite{Agakishiev:2010ur} for trigger particle in-plane and out-of-plane.  The solid lines represent the same subtraction procedure, but starting with only the first four Fourier harmonics of the unsubtracted data (the right side of Eq.~\eqref{dndphi}).  Note first that all useful information is contained in these first few harmonics.  

The fact that the double-peak structure on the away side depends on $\phi_s$ was taken as evidence that triangular flow does not represent a major contribution (since triangular flow should not depend on event plane orientation), and by extension that collective flow alone can not therefore explain the subtracted signal.  As we have seen, however, the odd Fourier harmonics indeed have no dependence on $\phi_s$, just as predicted, and therefore they are the same in the upper and lower panels of the figure.  All dependence on $\phi_s$ comes only from the even harmonics, which have only been partially subtracted by the ZYAM scheme.

It should be noted that this is not only because $v_2$ has been underestimated.  Consider the right column of Fig.~\ref{fig:ZYAM}.  Here, instead of using the reported values of $v_2$ and $v_4$, I simply use the actual extracted Fourier harmonics $v_2^{(a)} v_2^{(t,R)}\equiv V_{2\Delta}$, $v_4^{(a)} v_4^{(t,R)}\equiv V_{4\Delta}$.  
The subtracted signal is still not independent of trigger angle, even when the actual even harmonics are what have been ``subtracted''.  This, of course, is because the effects of the even harmonics are still present --- the subtraction scheme didn't actually remove them, because the ZYAM assumption is not true for the remaining odd harmonics.

A more thorough illustration can be found in Ref.~\cite{Ma:2010dv}.  There, the authors used the AMPT Monte Carlo model to investigate the dihadron correlations.  By randomizing the azimuthal angle of the transverse momenta of initial partons in their HIJING initial conditions, or turning off jet production completely, they can isolate the effect of jet correlations.  They indeed conclude that no non-flow correlations exist at large relative rapidity in their simulations (only effects from ``soft hot spots'', i.e., the initial geometry of the system.)  

More importantly their use of ZYAM subtraction illustrates several of these problems.  They are able to independently calculate all $v_n$ flow coefficients, and so have excellent control over the background used in this subtraction (and they include all coefficients, including triangular flow).   Nevertheless, because of the mentioned difficulties, they still find characteristic double-hump signals --- even when jet correlations are completely turned off (see, e.g., the blue dash-dot curve in their Fig.~5 \cite{Ma:2010dv}).

Even in the best of circumstances, then, the ZYAM background subtraction procedure can produce results that are difficult to analyze.  The characteristic shoulder structures are seen even with access to the best possible knowledge of flow coefficients and essentially no non-flow signal (i.e., only very small correlations from resonance decays), and it is difficult to differentiate between this case and when there are interesting non-flow signals present. 

A better way to analyze dihadron correlations is actually illustrated by STAR in Ref.~\cite{Agakishiev:2010ur}.  It could be argued that the most reliable way that we have to remove non-flow correlations is to introduce a large gap in pseudorapidity.  Therefore, we can simply take the long-range correlation (at large $\Delta\eta$) as our measurement of flow.  Then, this flow correlation can be directly subtracted to investigate the non-flow signals present at shorter ranges.  An example of this type of procedure was used in Ref.~\cite{Agakishiev:2010ur} to analyze ``jet-like'' correlations at $|\Delta\eta| < 0.7$.  This completely removes any assumption about the nature of non-flow correlations, except that is negligible at large relative pseudorapidity.  As we have seen, this is a well-supported assumption.
\section{Conclusions}
In conclusion, I have shown that long-range two-particle correlations are consistent with being entirely generated by collective flow.  Furthermore, analyzing such data with a ZYAM subtraction scheme can be misleading.  Any ZYAM-based analysis requires precise knowledge of the flow background and careful attention to effects of flow fluctuations, which show up in every Fourier component (not just triangular flow), along with a reliance on the zero-yield-at-minimum assumption.  I argue that it is better to take the long-range correlation as our best measurement of flow itself.   One can then subtract this from shorter-range correlations to study non-flow effects without any assumption about the nature of the non-flow signal 
  --- similar to what was done by STAR to study jet-like correlations in Ref.~\cite{Agakishiev:2010ur}.  
\begin{acknowledgments}
The author would like to thank J.-Y.\ Ollitrault for extensive discussions, and F.\ Wang for providing the experimental STAR data.  He would also like to thank the University of S\~ao Paulo, where much of this work was done, and the Funda\c c\~ao de Amparo \`a Pesquisa do Estado de S\~ao Paulo (FAPESP) for support during the visit.  
This work was funded by ``Agence Nationale de la Recherche'' under grant
ANR-08-BLAN-0093-01.
\end{acknowledgments}
\end{document}